\documentclass{aa}  

\usepackage{graphicx}
\usepackage{gensymb}
\usepackage{txfonts}
\usepackage{xspace}
\usepackage{hyperref}
\usepackage{xcolor}
\usepackage{academicons}
\usepackage{natbib}
\usepackage[normalem]{ulem}
\bibpunct{(}{)}{;}{a}{}{,} % to follow the A&A style

\newcommand{\blue}[1]{\textcolor{blue}{#1}}

\newcommand{\orcid}[1]{\href{https://orcid.org/#1}{\textcolor[HTML]{A6CE39}{\aiOrcid}}}

\newcommand{\srg}{\textit{SRG}\xspace}
\newcommand{\art}{{ART-XC}\xspace}
\newcommand{\srga}{{SRGA\,J144459.2$-$604207}\xspace}
\newcommand{\srgash}{{SRGA\,J1444}\xspace}
\newcommand{\flux}{erg\,cm$^{-2}$\,s$^{-1}$}

\newcommand{\be}{\begin{equation}}
\newcommand{\ee}{\end{equation}}
\newcommand{\bea}{\begin{eqnarray}}
\newcommand{\eea}{\end{eqnarray}}

\begin{document}

\title{{\it SRG}/ART-XC discovery of SRGA\,J144459.2$-$604207: \\ a well-tempered bursting accreting
millisecond X-ray pulsar}

\author{Sergey~V.~Molkov\inst{1} \and Alexander~A.~Lutovinov\inst{1} \and Sergey~S.~Tsygankov\inst{2} \and Valery~F.~Suleimanov\inst{3} \and Juri~Poutanen\inst{2} \and Igor~Yu.~Lapshov\inst{1} \and Ilia~A.~Mereminskiy\inst{1} \and Andrei~N.~Semena\inst{1} \and  Vadim~A.~Arefiev\inst{1}  \and Alexey~Yu.~Tkachenko\inst{1}}

\institute{
$^1$Space Research Institute, Russian Academy of Sciences, Profsoyuznaya 84/32, 117997 Moscow, Russia\\
$^2$Department of Physics and Astronomy,  FI-20014 University of Turku, Finland\\
$^3$Institut f\"ur Astronomie und Astrophysik, Universit\"at T\"ubingen, Sand 1, D-72076 T\"ubingen, Germany
}

%   \date{Received September 15, 1996; accepted March 16, 1997}

\abstract 
{We report on the discovery of the new accreting millisecond X-ray pulsar \srga\ using the {\it SRG}/\art\ data.  
The source was observed twice in February 2024 during the declining phase of the outburst.
Timing analysis revealed a coherent signal near 447.8~Hz modulated by the Doppler effect due to the orbital motion. 
The derived parameters for the binary system are consistent with the circular orbit with a period of $\sim5.2$~h.  
The pulse profiles of the persistent emission, showing a sine-like part during half a period with a plateau in
between, can well be modelled by emission from two circular spots partially eclipsed by the accretion disk. 
Additionally, during our 133~ks exposure observations, we detected 19 thermonuclear X-ray bursts. 
All bursts have similar shapes and energetics, and do not show any signs of photospheric radius expansion. 
The burst rate decreases linearly from one per $\sim$1.6~h at the beginning of observations to one
per $\sim$2.2~h at the end and anticorrelates with the persistent flux. 
Spectral evolution during the bursts is consistent with the models of the neutron star atmospheres heated by
accretion and imply a neutron star radius of 11--12~km and the distance to the source of 8--9~kpc.
We also detected pulsations during the bursts and showed that the pulse profiles differ substantially from
those observed in the persistent emission. 
However, we could not find a simple physical model explaining the pulse profiles detected during the bursts. 

}

    
\titlerunning{{\it SRG}/ART-XC discovery of SRGA\,J144459.2$-$604207}
\authorrunning{Molkov et al.} 

\maketitle
%
%-------------------------------------------------------------------
\section{Introduction}

Accreting millisecond X-ray pulsars (AMXPs) constitute a relatively small group of binary systems featuring
a rapidly rotating neutron star (NS) and a low-mass optical companion
\citep[see][for recent reviews]{patruno21,disalvo22review}. The NSs in these systems, believed to be
progenitors of rotation-powered millisecond pulsars, exhibit spin periods ranging from $\sim1.7$ to
$\sim10$~ms and possess relatively weak magnetic fields (around $10^8$--$10^9$~G). Thus these object
 play an important role in the star evolutionary processes, but the current sample of known AMXPs comprises
only about two dozen sources. Therefore search for such objects and their discoveries are quite important task.
Moreover this task is very non-trivial one and demands extraordinary technical capabilities of X-ray
instruments, including a large effective area and high time resolution. 

The {\it Mikhail Pavlinsky} \art\, telescope \citep{2021A&A...650A..42P} onboard
the {\it Spectrum-Roentgen-Gamma} observatory (\srg; \citealt{sunyaev21}) discovered a new AMXP, \srga
(hereafter SRGA\,J1444), during the ongoing all-sky survey. 
The source was found on 2024 Feb 21 at the position close to the Galactic plane with
coordinates ($l,b$) = ($316\fdg4, -0\fdg8$) and the flux of $\sim 100$~mCrab in the 4--12 keV energy
band \citep{ATel16464}. 

The intense follow-up campaign carried out immediately after the discovery revealed that \srgash is a new
accreting millisecond pulsar with a spin period of 447.8~Hz \citep{2024ATel16474....1N} showing regular
Type-I X-ray bursts \citep{2024ATel16475....1M,2024ATel16480....1R,2024ATel16485....1S}. 
Subsequent observations of \srgash with NICER and {\it Insight-HXMT} instruments unveiled a clear sinusoidal
Doppler shift of the spin frequency that allowed to determine an orbital period of $\sim$5.2~h, indicating
a companion star mass exceeding 0.255 M$_\odot$  \citep{2024ATel16480....1R, 2024ATel16548....1L}.

The improved coordinates of \srgash, RA (J2000) = 14$^{\rm h}$44$^{\rm m}$58\fs9, Dec (J2000) = $-$60\degr41\arcmin55\farcs3, were obtained with the High Resolution Camera on board the {\it Chandra} observatory \citep{2024ATel16510....1I}. Despite the accurate localization of the source, no optical/IR counterparts were found \citep{2024ATel16476....1S, 2024ATel16477....1C, 2024ATel16487....1B, 2024ATel16489....1S}. A radio counterpart was discovered at the position of \srgash using the Australia Telescope Compact Array \citep{2024ATel16511....1R}. Its spectral index consistent with the emission from either a compact jet or discrete ejecta from an X-ray binary. 

Interestingly, that the retrospective analysis of the MAXI and {\it INTEGRAL} archival data revealed past X-ray activity of \srgash. Particularly, an increase in the flux from the sky position coincident with the \srgash one was observed in the beginning of Jan 2022 and mid-Dec 2023 \citep{2024ATel16483....1N, 2024ATel16493....1S}.

In this paper, we report on the discovery of a new AMXP \srgash using the \art telescope data. Results of timing and spectral analysis both the persistent emission and its evolution during multiple Type-I bursts are presented as well as modelling of the bursts and pulse profile.

%--------------------------------------------------------------------

\section{Observations and data reduction}
\label{sec:observations}

%=======================================
\begin{figure}
%\centerline{\includegraphics[width=\columnwidth]{figs/srga_maxi_art_lc.pdf}}
\centerline{\includegraphics[width=\columnwidth]{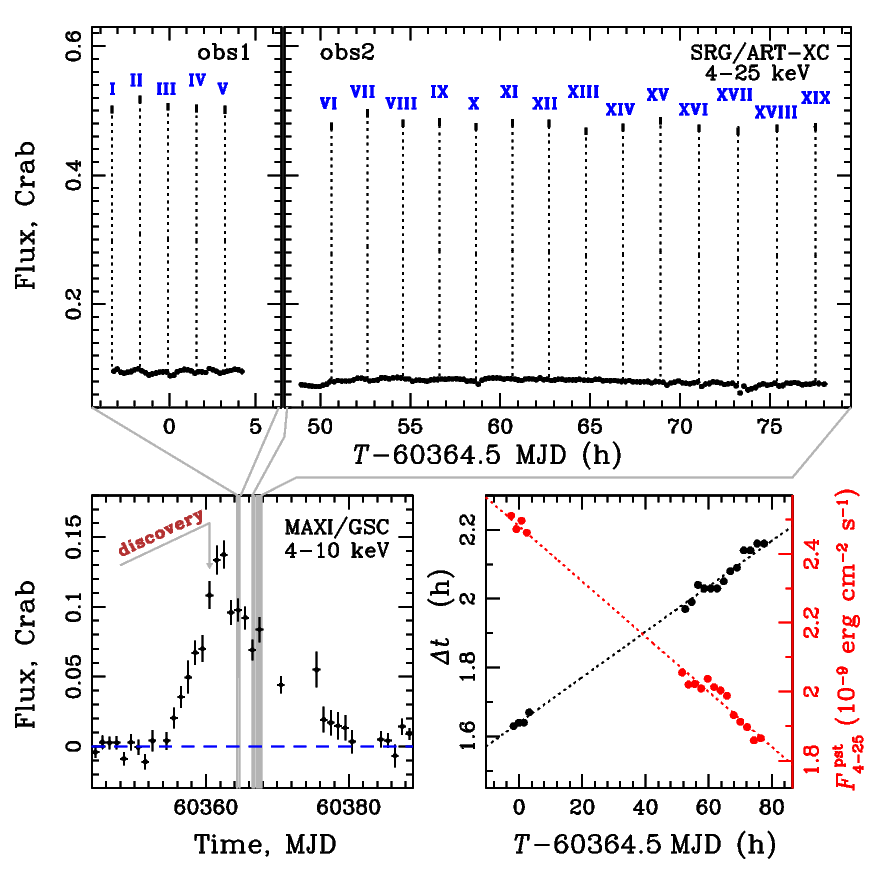}}
\caption{Temporal evolution of the \srgash\ emission during the 2024 outburst. 
Upper panel: the \art light curve during two observations. For type-I X-ray bursts
their time and mean flux are shown with the dashed line and thick streak. 
Bottom left: The flux evolution as observed by MAXI/GSC in the 4--10~keV energy band.
Bottom right: the X-ray burst recurrence time (in black, left axis) and the mean
persistent flux between the bursts in the 4--25~keV energy band $F_{\rm 4-25}^{\rm pst}$
(in red, right axis) as measured by \art.}
\label{fig:overall_lc}
\end{figure}
%=======================================

\srgash was first detected in the data downlinked from the \art telescope on  2024 Feb 21 using a near-real-time data processing chain. 
It was quickly identified as a new X-ray source located in the Galactic plane.
At the time of the detection the source flux was about $(1.9\pm0.1)\times10^{-9}$ \flux in the 4--12 keV band \citep[][see also Fig.~\ref{fig:overall_lc}]{ATel16464}.
Two extended observations were conducted with the \art telescope as part of the follow-up program. The first observation began on 2024 Feb 2  and lasted for 28~ks, while the second observation started on 2024 Feb 26 and continued for 105~ks.

The \art telescope is an imaging instrument consisting of seven modules and operating in the photon counting mode \citep{2021A&A...650A..42P} in the 4--30 keV energy band. The data were processed using the {\sc artproducts} v1.0 software with the latest CALDB {\sc v20220908}.
For both spectral and timing analysis we extracted photons from the circle of 1\farcm8 radius around the source position applying also an energy filtering and excluding high energies ($>25$ keV).   
Due to the limited energy band of \art, which does not permit a reliable determination of a moderate interstellar absorption, for energy spectra fitting we employed models with the fixed equivalent hydrogen column density of $2.7 \times 10^{22}$~cm$^{-2}$, determined from  the NICER data \citep{2024ATel16474....1N}. All reported flux values in the paper are unabsorbed unless explicitly stated otherwise.

\section{Persistent emission and bursting behavior}\label{sec:pers_emi}

%=======================================
\begin{figure}
\centerline{\includegraphics[width=\columnwidth]{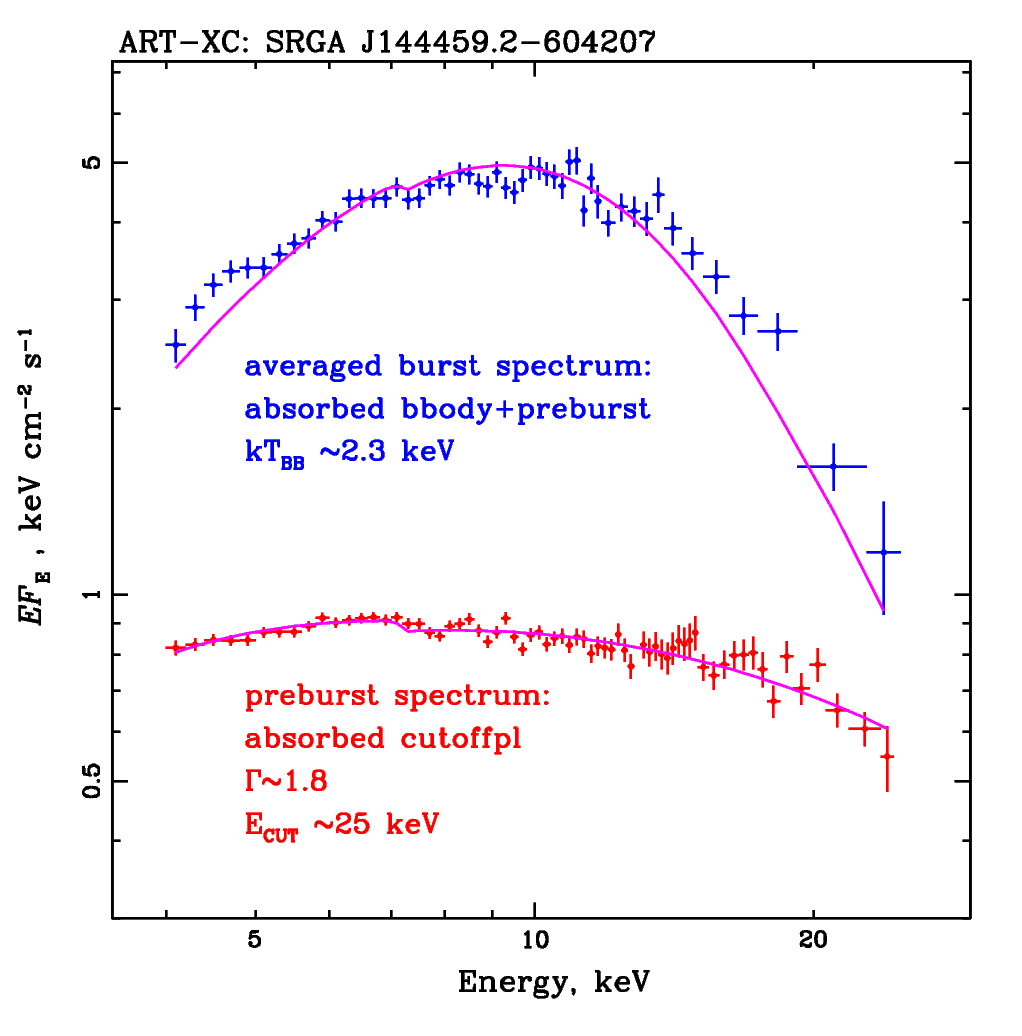}} 
\caption{Average energy spectrum of all type-I X-ray bursts observed
by \art\ during first observation (blue crosses) and the spectrum of the persistent
emission extracted between bursts number II and III (red crosses). 
The pink curves represent the spectral model of the absorbed blackbody and the cutoff
power law, respectively. }
\label{fig:average_specs}
\end{figure}
%=======================================

Observations of \srgash with the \art telescope were carried out soon after the maximum of the outburst at the declining phase (Fig.~\ref{fig:overall_lc}, lower left panel). In total, in two observations, a total of 19 Type-I X-ray bursts were registered, each lasting about 45~s and occurring at approximately equal intervals of about two hours. However, a detailed study showed that the recurrent time $\Delta t$ between bursts is not constant, and the burst rate decreases linearly from one per $\sim$1.6~h at the beginning of observations to one per $\sim$2.2~h at the end (Fig.~\ref{fig:overall_lc}, lower right panel), while the energy release during the bursts remains approximately at the same level (Fig.~\ref{fig:overall_lc}, upper panel). 

To trace the evolution of the persistent emission, we split our observations into intervals of about 800~s, with X-ray bursts being excluded from the analysis. 
Then we determined the average count rate in the 4--25~keV energy band for each interval. 
The resulting light curve in Crab flux units is presented on the upper panel of Fig.~\ref{fig:overall_lc}.

To quantify the rate of change in the persistent flux and establish its relationship with the burst generation frequency, we first reconstructed the energy spectra during intervals between the bursts. Then we fitted these spectra with a simple analytical model consisting of the absorbed power law with a high energy cutoff, \texttt{tbabs$\times$cutoffpl} in \textsc{xspec} \citep{Arnaud96}, and estimated the corresponding fluxes. All spectra have a similar shape and are well approximated by this model with the photon index of $\Gamma \sim 1.8$ and high energy cutoff energy $E_{\rm cut}\sim25$~keV and differ only by their normalizations. As an example, one of the spectra of the persistent emission extracted between the bursts II and III is shown in Fig.~\ref{fig:average_specs}. 
The evolution of the flux between the bursts is presented on the lower right panel in Fig.~\ref{fig:overall_lc} (in red) and it is clearly seen that this flux is anticorrelated with the bursts recurrence time. At the same time, the total energy released between bursts in the 4--25~keV energy band remains approximately constant, with the fluence being 
$E_{\rm bet\_brst} \simeq1.48\times10^{-5}$~erg~cm$^{-2}$.

\subsection{Type-I X-ray bursts}\label{sec:brsts}

%=======================================
\begin{figure}
\centerline{\includegraphics[width=0.85\columnwidth]{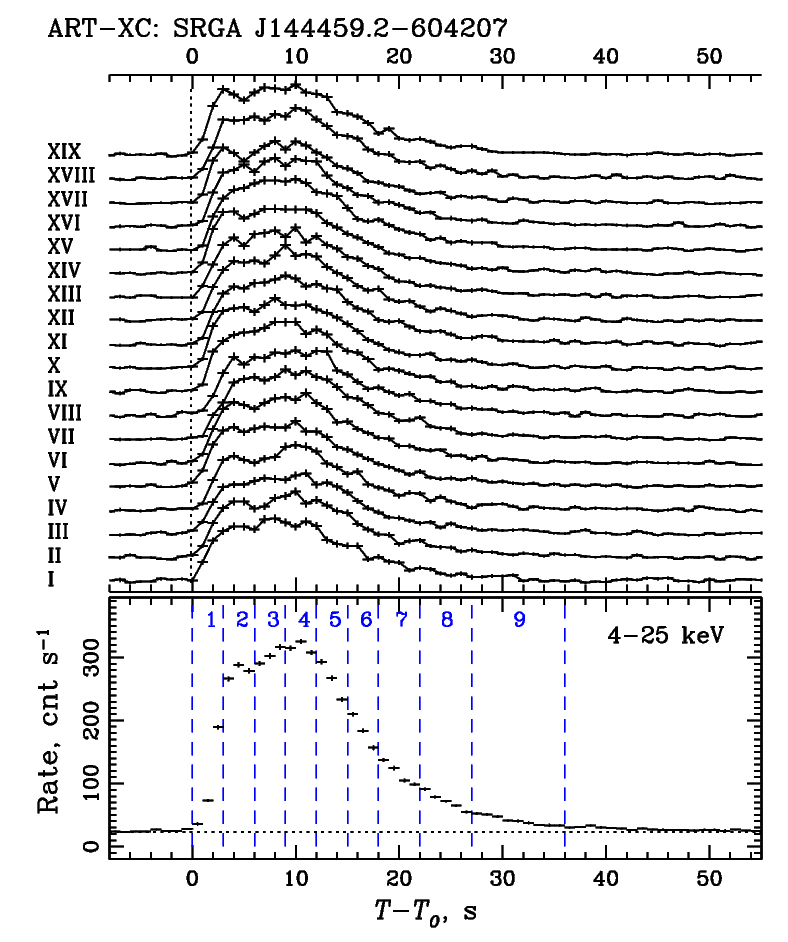}}
\caption{Light curves of all detected type-I X-ray bursts in the 4--25~keV energy band. 
The bottom panel shows the averaged burst profile. 
Dashed vertical lines represent the time intervals used for the time-resolved spectral analysis
(see Sect.~\ref{sec:brsts_spectroscopy}).
}
\label{fig:all_burst_shape}
\end{figure}
%=======================================

First of all, the averaged energy spectrum for each of the bursts was analysed. 
We reconstructed \art spectra in the 4--25~keV energy band collected during 45~s time intervals starting from the burst beginning. Note, that the total spectrum of the source during the burst consists of the persistent emission and the spectrum of a thermonuclear flash itself. 
The contribution of the persistent emission to the total burst spectrum was estimated by constructing the energy spectrum of the source during 800~s preceding the burst. All pre-burst spectra are well approximated by the cut-off power-law model and the same parameters within the uncertainties, as for the persistent emission (see Sect.~\ref{sec:pers_emi}). In order to describe the energy spectra of thermonuclear bursts, we approximated them with the two-components model, including a blackbody and a cut-off power law  (modified by interstellar absorption with fixed equivalent hydrogen column density, see above). The parameters of the second component, excluding normalization, were fixed at the values obtained for the persistent emission. 

This model describes all 19 bursts relatively well with the same  within errors blackbody flux of $F_{4-25}\simeq 8\times10^{-9}$~erg~s$^{-1}$~cm$^{-2}$ and the temperature, $kT_{\rm bb}\simeq 2.3$~keV (see Fig.~\ref{fig:average_specs}). 
The fluence released per burst in the 4--25~keV energy band is
about $E_{4-25}^{\rm bst}\simeq 3.6\times10^{-7}$~erg~cm$^{-2}$. 

We investigated also morphology of the bursts and found that all the bursts have a similar shape with the fast rise that transforms into a plateau-like phase which in turn changes to an exponential decay (see upper part of Fig.~\ref{fig:all_burst_shape}).
The similarity of the burst profiles and their spectral parameters gives us the opportunity to analyze not each burst individually, but their sum, which significantly improves statistical errors and allows us to study emission properties with the better time resolution.
In the lower panel of Fig.~\ref{fig:all_burst_shape}, we present an average burst profile with a time resolution of 1~s. 
Below, we will take a closer look at the spectral evolution during the bursts.

\subsection{Time resolved  spectral analysis of the bursts}\label{sec:brsts_spectroscopy}

%=======================================
\begin{figure}
\centerline{\includegraphics[width=0.85\columnwidth]{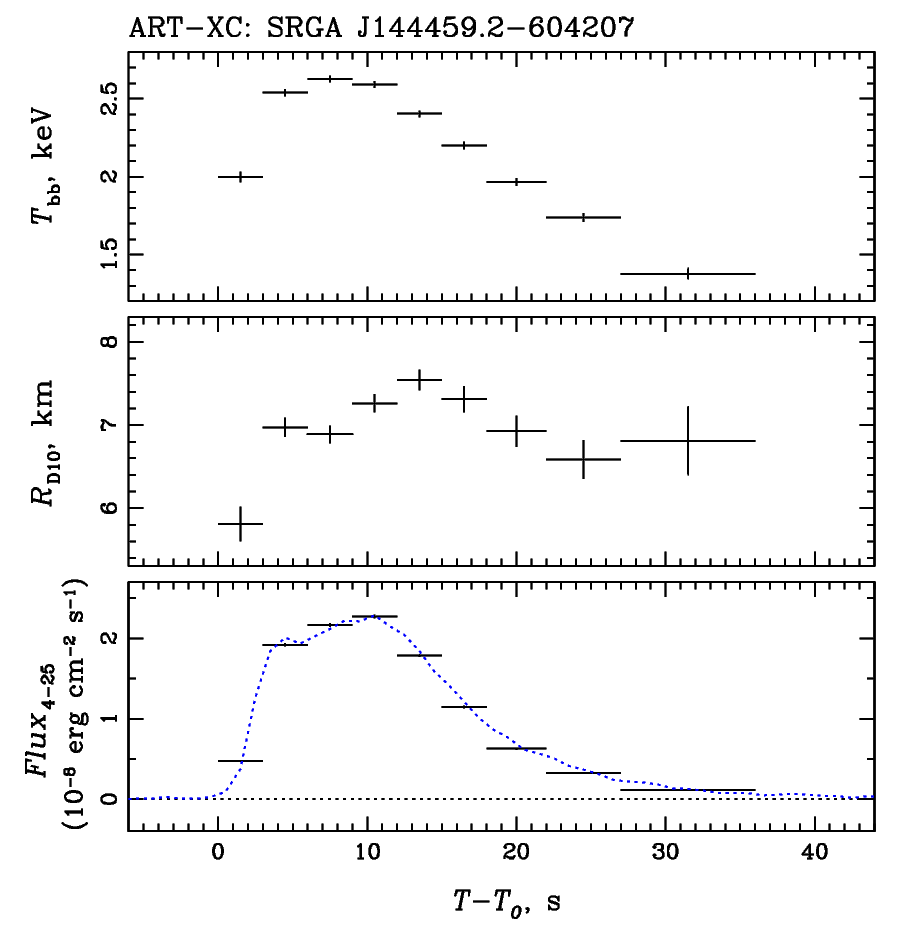} }
\caption{Time resolved spectroscopy of the averaged X-ray bursts emission. 
The blue dotted line on the bottom panel shows the count rate light curve in the 4--25 keV with subtracted pre-burst rate and renormalized in amplitude. 
}
\label{fig:average_burst_params}
\end{figure}
%=======================================

In order to investigate the spectral evolution of the radiation during X-ray bursts, we divided the averaged burst profile into 9 time intervals (see the bottom panel of Fig.~\ref{fig:all_burst_shape}) and reconstructed energy spectra for them. 
The contribution of the persistent emission was taken into account in the same way as described above, i.e. using it as background for approximating the bursts spectra. To describe these spectra, we used the blackbody model (modified by the fixed interstellar absorption), which adequately describes the spectra for all time intervals. 
The evolution of the blackbody  parameters during the bursts is shown in Fig.~\ref{fig:average_burst_params}.
The blackbody normalization $K_{\rm bb}$ was converted to the radius of the emission region for the distance of 10~kpc: $R_{\rm D10}=\sqrt{K_{\rm bb}}$. 
Also, in the lower part of the figure, we showed the flux evolution in the energy range of 4--25~keV. Note that there are no signatures of a photospheric radius expansion during the bursts. 
The burst peak bolometric luminosity, $L_{\rm pk}=3.2\times10^{38}$~erg~s$^{-1}$ 
is reached at about tenth second, and the total energy release during the burst is approximately $E_{\rm bst}=5.1\times10^{39}$~erg (again the 10~kpc distance is assumed).

\section{Timing properties}\label{sec:timing}

In this section we investigate timing properties of \srgash, including pulsations, orbital motion as well as an evolution of pulse profile with the time and energy both for the persistent emission and during the type-I X-ray bursts. 

%=======================================
\begin{figure}
\centerline{\includegraphics[width=0.85\columnwidth]{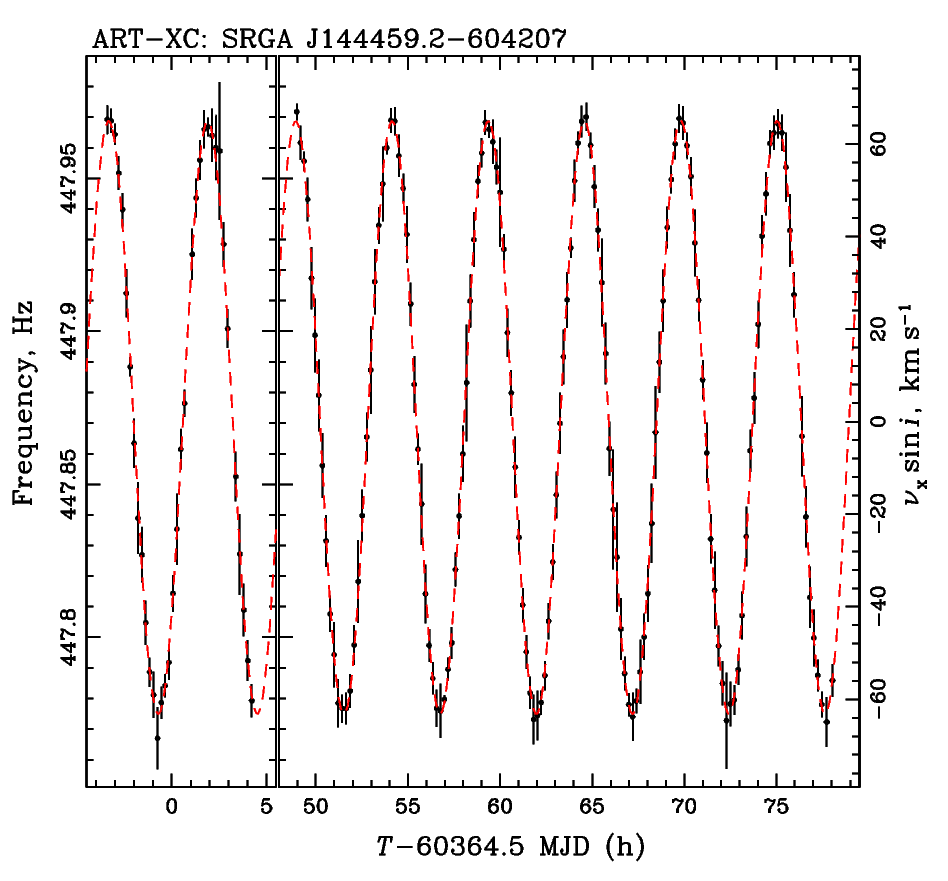}}
\caption{Variations of the measured frequency of the NS rotation due to an orbital motion in the binary system according to the \art data (black points). The model is shown with the red dashed-line. }
\label{fig:freq_evol}
\end{figure}
%=======================================

\subsection{Orbital parameters and coherent timing}\label{sec:coher_timing}

A coherent signal near 447.8~Hz from \srgash was discovered by NICER \citep{2024ATel16474....1N} and later parameters of the binary motion had been preliminary determined based on the long set of the NICER observations \citep{2024ATel16480....1R}. Pulsations along with the sinusoidal modulation of its frequency with time are also clearly detected in the \art data. 
To determine the spin period and orbital parameters, our observations were divided into segments of approximately 800~s for the subsequent analysis (type-I X-ray bursts were excluded). Power spectra were then computed for each data segment to measure the frequency associated with this spin period.
The resulting spin-frequency evolution is presented in Fig~\ref{fig:freq_evol}. 
Further, assuming a circular motion, we obtain the following orbital parameters for the binary system: $P_{\rm orb}=0.217649(5)$~d is the orbital period, the projected semi-major axis is $a_{\rm x} \sin i = 0.6513(2)$~lt-s and $T_{\rm asc}=$MJD~60361.64126(5) is the time of passage through the ascending node.
These values are in good agreement with the parameters obtained from the NICER  \citep{2024ATel16480....1R} and {\it Insight-HXMT} \citep{2024ATel16548....1L} data. 
Subsequently, we corrected the arrival times of photons for the orbital motion using the obtained orbital parameters. Following this correction, we re-evaluated the NS rotation period for each of the time segments.

\begin{figure}
\centerline{\includegraphics[width=0.85\columnwidth]{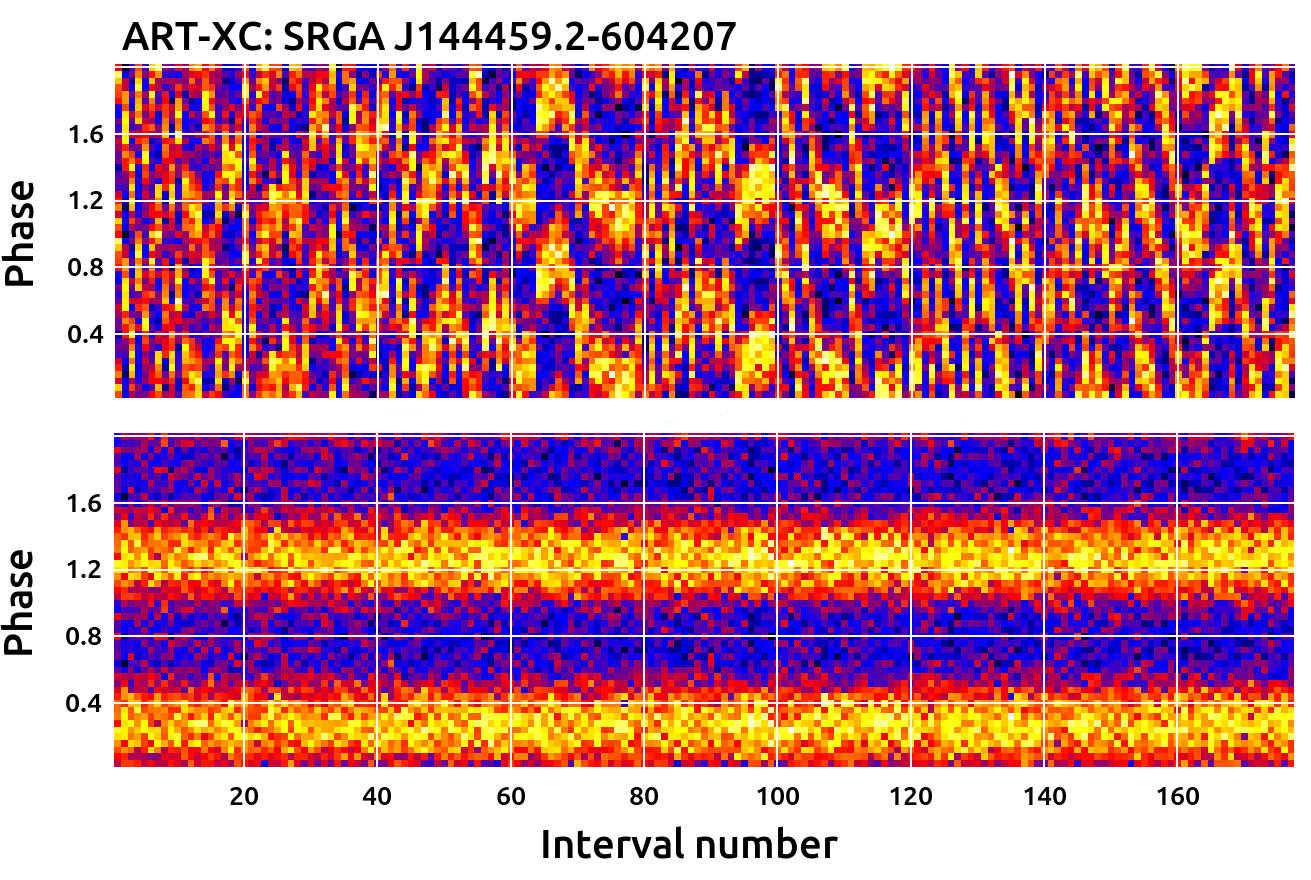}}
\caption{Upper panel: Light curves for each of segments folded with period determined for this interval assuming the same epoch time ($T_{\rm epoch}=$MJD~60361.0).
Bottom panel: Phase aligned pulse profiles for all segments.}
\label{fig:ph_alignment}
\end{figure}

We found that the measured spin period evolves with time in a complex way, which may indicate both that our
orbital solution is not accurate enough and that the period does change, for example, due to variations
in the mass accretion rate. 
At first, we folded the data for each time segment with the corresponding spin period with the same
epoch time ($T_{\rm epoch}=$MJD 60361.0). 
The resulting segment-by-segment folding is presented on the upper panel of Fig.~\ref{fig:ph_alignment}. 
We see that the pulsations are reliably detected but their phase floats from segment to segment. 
For further phase-resolved analysis, it is necessary either to build a more accurate model of orbital motion
and take into account possible intrinsic spin variations in time, or to use the existing model for
individual segments, but simply introduce a correction for phase shift for each segment. 
The former, i.e. the proper phase-connected analysis, is much more complicated and, among other things,
implies a well-calibrated or well-modeled on-board clock for the entire observation time, which in our
case has not yet been done. 
The second approach is much simpler, but at the same time allows to do a full-fledged phase-resolved
analysis.
Therefore, we followed the second way and attributed to each segment not only the corresponding period,
but also the phase shift necessary for phase alignment. 
To determine the shifts we folded the data for each interval with the corresponding period starting from
the epoch time ($T_{\rm epoch}=$MJD 60361.0) and approximated the resulted profile with the simple sine model.
Using the phase values at which the sine turns to zero, we aligned the solutions for the intervals so that
the sine equals a zero value at phase zero.
Folding the data using this ``phase-aligned model'' is shown in the bottom panel of Fig.~\ref{fig:ph_alignment}. 
Below, this model will be used for the reconstruction of pulse profiles for the persistent emission.

%=======================================
%=======================================

%=======================================
\begin{figure}
\centerline{\includegraphics[width=0.85\columnwidth]{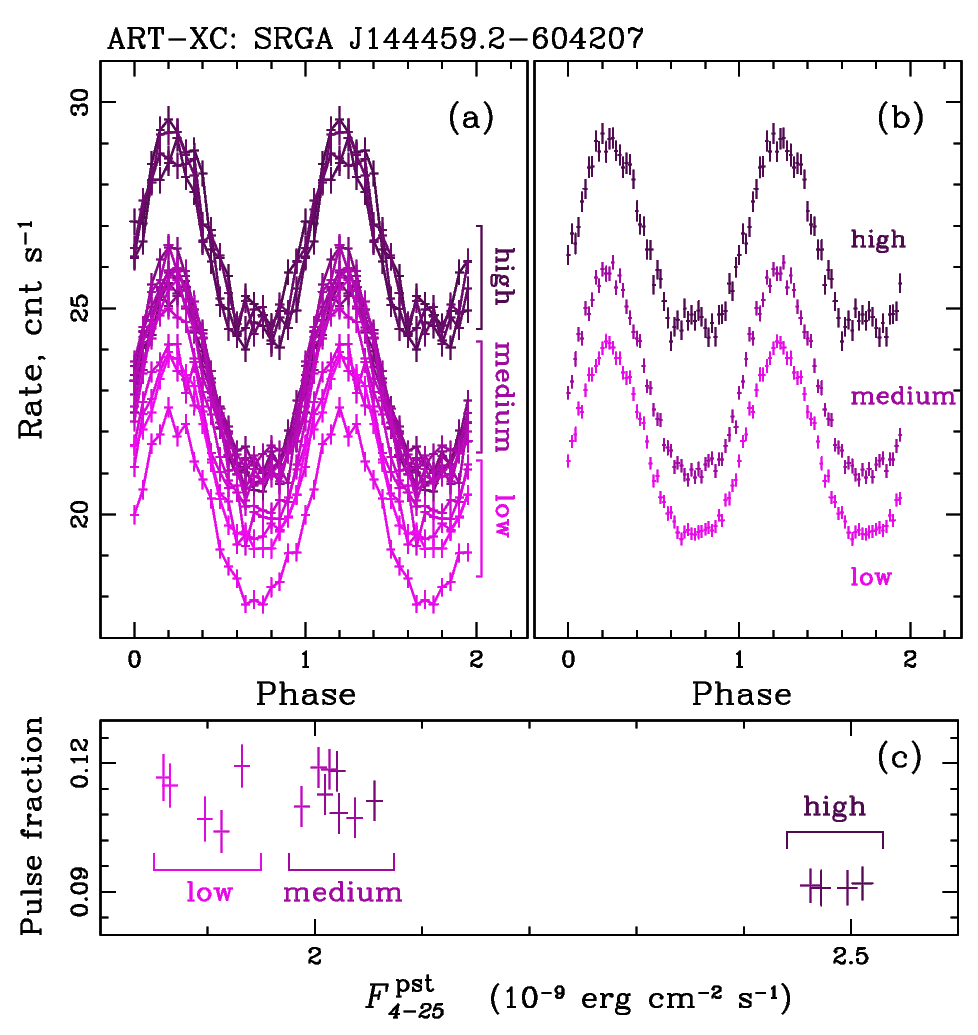} }
\caption{Evolution of the pulse profile (panels a and b) and pulsed fraction (panel c) with the source intensity.}
\label{fig:p_prof_and_PF}
\end{figure}
%=======================================

\subsection{Pulse profile evolution}

\subsubsection{Pulse profiles of the persistent emission}

As the first step, we investigate how the pulse profile is changing with the flux for the persistent emission. To do this, we used 17 time intervals between successive type-I X-ray bursts and folded the 4--25~keV source light curve using the phase-aligned model (see Sect.~\ref{sec:coher_timing}). 
The folded curves are shown in Fig.~\ref{fig:p_prof_and_PF}a. Using a standard expression of $PF=(C_{\max}-C_{\min})/ (C_{\max}+C_{\min})$, where $C$ is a bin count rate, we estimated the pulsed
fraction for each of the intervals and traced its value with the flux.
The dependence of the pulsed fraction on the flux in the energy range of 4--25~keV is shown in  Fig.~\ref{fig:p_prof_and_PF}c. We see that the PF is consistent with being constant at low and medium fluxes and decreases at high fluxes. It follows from Fig.~\ref{fig:p_prof_and_PF}a that the pulse profile in a wide energy range have only minor difference. Particularly, at lower fluxes, it can be described by a sine wave at phases 0.0--0.5 and a plateau in the range of 0.5--1.0. At higher fluxes, a small bump (interpulse) appears instead of the plateau. To demonstrate the difference more clearly, we have made pulse profiles for three flux levels (Fig.~\ref{fig:p_prof_and_PF}b).
The smaller pulsating fraction in ``high state'' is probably due to the presence of the interpulse.

\begin{figure}
\centerline{\includegraphics[width=0.9\columnwidth]{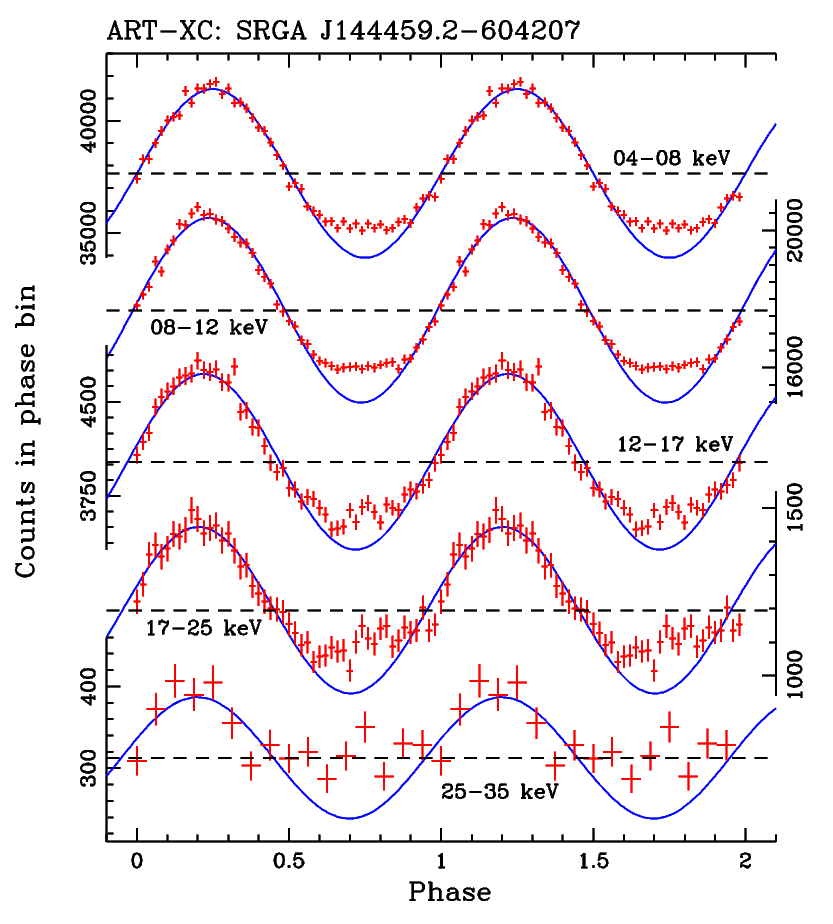}}
\caption{Pulse profiles of the persistent emission from the source in five energy bands. 
Blue lines are the approximations with a sine wave in the phase interval 0.0--0.5.}
\label{fig:pulse_prof_5en}
\end{figure}

At the next step, an evolution of the pulsed emission with the energy was investigated.
We divided the 4--25~keV energy range of the \art telescope into 4 bands (4--8, 8--12, 12--17, and 17--25~keV) and reconstructed light curves for all observations in each band, excluding data intervals containing X-ray bursts. We also considered an additional channel of 25--35~keV (technically the telescope mirrors can focus photons with the energy of up to 35~keV). In a normal situation, this range is not used due to the low efficiency of the events registration and the difficulty with absolute flux calibrations, however, for qualitative analysis, it can be quite representative. 
Resulting pulse profiles in different energy bands are presented in Fig.~\ref{fig:pulse_prof_5en}. 
It is clearly seen that the pulse shape depends weakly on the energy. Additionally, in all energy ranges, the ``sine-like'' wave part is present, while the shape of the ``plateau'' is varied.  
There is also a small but noticeable phase shift $\Delta\phi$ with energy, with the harder photons arriving earlier than the softer ones, e.g., $\Delta\phi=-0.05$ for the 17--25~keV photons relative to 4--8 keV photons (i.e. the slope of the linear relation of $\Delta\phi(E)$ is $-0.033$~keV$^{-1}$).   

%=======================================
%=======================================

\subsubsection{Pulse profiles during X-ray bursts}

\begin{figure}
\centerline{\includegraphics[width=0.85\columnwidth]{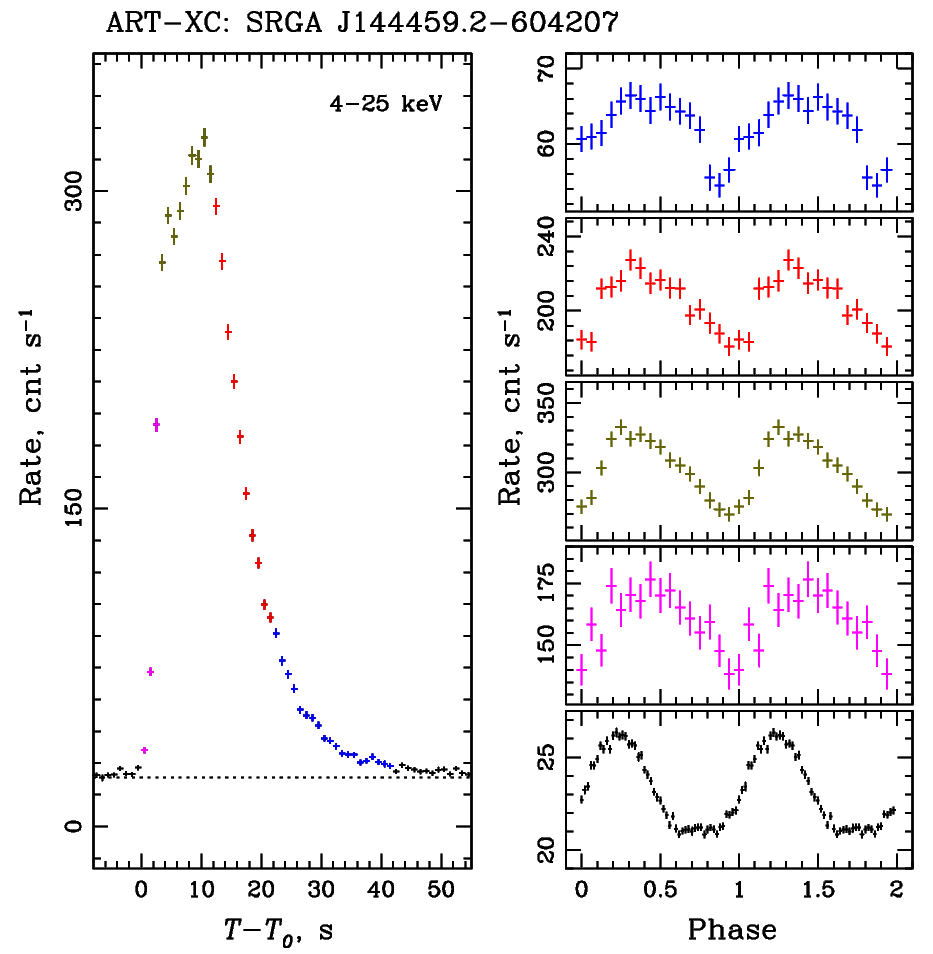} }
\caption{Pulse profile evolution during type-I X-ray bursts. 
Appropriate time intervals are coded by color. 
The pulse profile of the persistent emission is shown in black at the bottom right panel. }
\label{fig:pulse_evol_dur_bursts}
\end{figure}

To track the possible evolution of the pulse profile during the bursts, we divided the bursts into 4 time intervals counting from the beginning of each burst: 0--3, 3--12, 12--18 and 18--36~s. The left part of Fig.~\ref{fig:pulse_evol_dur_bursts} shows the light curve of the source averaged over all X-ray bursts and the selected intervals are highlighted by different colors. Results of applying the folding procedure to the light curves extracted from these time intervals are shown on the right panels of Fig.~\ref{fig:pulse_evol_dur_bursts}. In addition, we also provide the folded light curve of the persistent emission of intervals of 1000~s before and after the bursts (the bottom right panel). The pulse profiles practically do not change during the burst and have a sinusoidal shape with a steeper left wing. This shape differs significantly from the shape of the pulse profile of the  persistent emission. The fraction of pulsating emission in the 4--25~keV energy range remains constant during the bursts, about 11\%, and is similar to the value of the pulsed fraction for persistent emission in the same energy band. 

%=======================================
%=======================================

\section{Discussion}
\label{sec:disc}

We have reported the discovery of \srgash, a new member of a small class of AMXPs.
We clearly detected pulsations from the source at a frequency of about 447.8~Hz and determined binary ephemeris based on the evolution of this frequency over time. 
Our solution is in good agreement with the ephemerides obtained from the data of NICER \citep{2024ATel16480....1R} and {\it Insight-HXMT} \citep{2024ATel16548....1L} observatories.
Based on the ephemeris, it is possible to obtain the pulsar mass function $f_{\rm x}\simeq0.00626~M_\odot$, which, in turn, gives an estimate for the mass of a normal companion in this system, $M_2>0.25~M_\odot$ (assuming the NS mass of $M_{\rm NS}=1.4~M_\odot$).

\subsection{Constraints from the burst/persistent X-ray emission}

Based on the observational values, we can estimate some parameters of the binary system or accretion flow. In particular, using the peak
burst luminosity value it is possible to estimate the distance to the system. Indeed, if we take the empirical value of the Eddington luminosity
$(2.2-3.8)\times 10^{38}$~erg~s$^{-1}$ that depend only on the hydrogen abundance in the burst fuel \citep{kuulkers03}
and the burst peak bolometric flux $F_{\rm pk}=2.67\times10^{-8}$~erg~cm$^{-2}$ s$^{-1}$, then an upper limit on the distance to the system will be $D<8.3-10.9$~kpc (a more advanced approach is presented in the next section).

%=======================================
\begin{figure}
\centerline{\includegraphics[width=\columnwidth]{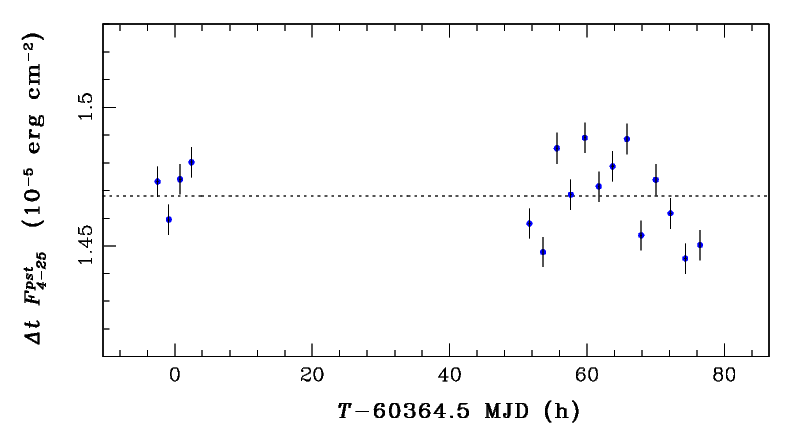} }
\caption{Dependence of the product ${\Delta t F_{4-25}^{\rm pst}}$ from time for 17 pairs of consecutive bursts.}
\label{fig:delta_fpst}
\end{figure}
%=======================================

Our data provides an opportunity to evaluate another key parameter, $\alpha$- ratio of accretion to thermonuclear energy, which allows us to evaluate the composition of bursting fuel, namely the mean hydrogen fraction at ignition, $\overline{X}$.
In other words, $\alpha$ shows how much the efficiency of energy release during accretion process is higher than in a thermonuclear reaction and can be calculated as a ratio of the energy released between two consecutive bursts to the energy released per burst.
In observational terms, this parameter can be expressed as follows:
\be
\label{alpha_est}
\alpha = \frac{\Delta t\, F_{4-25}^{\rm pst} C_{\rm bol}}{E_{\rm b}},
\ee
where $\Delta t$ is the burst recurrence time, $F_{4-25}^{\rm pst}$ is the mean persistent flux between bursts, $C_{\rm bol}$ is the coefficient to convert the persistent flux from a narrow energy band to bolometric flux and $E_{\rm b}$ is the burst fluence
\citep[see ][and references there for more details]{galloway22}.
We have 17 pairs of consecutive bursts by which we can estimate the product ${\Delta t\ F_{4-25}^{\rm pst}}$, and all products within the error range take on the value $1.47\times 10^{-5}$~erg~cm$^{-2}$ (Fig.~\ref{fig:delta_fpst}).
We estimated $C_{\rm bol}$ using the model describing the spectrum of persistent emission, extended the energy range to 0.2--60~keV (in which the vast majority of energy is released) and according to the model the flux in this energy range is about 3 times higher than the flux in the range 4--25~keV. 
Bolometric fluence $E_{\rm b}=4.18\times10^{-7}$~erg~cm$^{-2}$ we calculated based on time resolved spectroscopy of the burst.
Substituting the values obtained above into Eq.~\eqref{alpha_est} we get $\alpha\simeq105$. 
From the measured value of $\alpha$ we can estimate hydrogen mass fraction $\overline{X}$ in the ig~nition layer from Eq.~(11) in \citet{galloway22}.
We get $\overline{X} \approx 0.16$ assuming the NS radius $R_{\rm NS}=11.2$~km and mass $M_{\rm NS}=1.4 M_\odot$, indicating that the fuel is helium-rich.

\subsection{X-ray bursts spectral evolution}

%=======================================
\begin{figure}
\centerline{\includegraphics[width=0.85\columnwidth]{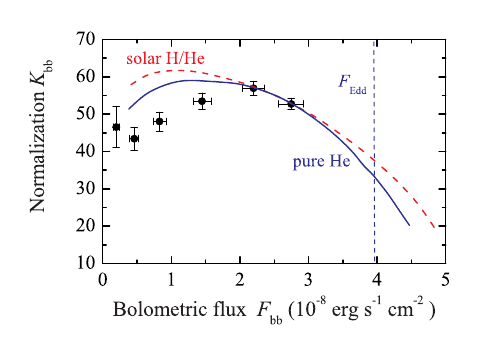} }
\caption{Evolution of spectral parameters of the averaged X-ray burst from \srgash. 
The best-fit model curves for solar H/He mix (red dashed curve) and pure He atmospheres (blue solid curve) are shown. 
The vertical dashed line correspond to the Eddington flux for pure He atmosphere model. }
\label{fig:Kbb-Fbb_model}
\end{figure}
%=======================================

Spectra of X-ray bursting NSs are well fitted by a diluted blackbody and can be described by the model spectra of hot NS atmospheres \citep{London86,Lewin93,Suleimanov12}.  
Spectral evolution of X-ray bursts which occurred during the hard persistent spectral state can be described with the sequences of model atmosphere spectra of decreasing relative luminosities \citep{Suleimanov12,Kajava14}.  
At the late burst stages, the atmospheres of the X-ray bursting NSs can be heated by renewed accretion which lead to certain changes in the emergent spectrum \citep{Suleimanov18}. 

Model hot NS atmosphere spectra can be approximated by a diluted blackbody spectrum \citep[see, e.g.,][]{Suleimanov11}
\be
    F_E = w\,\pi B_E (f_{\rm c} T_{\rm eff}),
\ee
where the effective temperature $T_{\rm eff}$ is the parameter of the model atmosphere, $w$ is the dilution factor, and $f_{\rm c}$ is the color correction factor.
Values of $w$ and $f_{\rm c}$ depend on $T_{\rm eff}$, surface gravity, and chemical composition of the atmosphere. 
Extended tables of $w$ and $f_{\rm c}$ for three chemical compositions (pure He, solar abundance, and solar H/He mix with the metal abundances reduced by a factor of 100), nine surface gravities, and more than twenty relative luminosities, were computed using the model atmospheres by \citet{Suleimanov12}. 
The effective temperature for  a given relative luminosity $\ell=L/L_{\rm Edd}$ depends on the surface gravity and the chemical composition.
The tables of $w$ and $f_{\rm c}$ can be used for interpretation of the X-ray burst spectral evolution  after the maximum flux if the observed spectra are fitted by the blackbody. 
%=======================================
\begin{figure}
\centerline{\includegraphics[width=0.85\columnwidth]{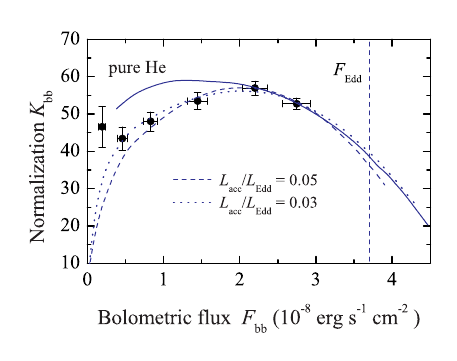} }
\caption{Same as Fig.~\ref{fig:Kbb-Fbb_model}, but showing the best-fits to the data with the accretion-heated pure He  atmosphere models of various accretion luminosities $L_{\rm acc}/L_{\rm Edd}$=0 (solid line), 0.03 (dotted), and 0.05 (dashed)  \citep{Suleimanov17}.  
The vertical dashed line gives  the Eddington flux for the model with $L_{\rm acc}/L_{\rm Edd}$=0.03. }
\label{fig:model_acc}
\end{figure}
%=======================================

From the parameters of the blackbody fits shown in Fig.~\ref{fig:average_burst_params}, we construct a dependence of the blackbody normalization $K_{\rm bb}$ (in units (km/10\,kpc)$^2$) on  the bolometric flux $F_{\rm bb}$ (in erg\,s$^{-1}$\,cm$^{-2}$) (see Figs.~\ref{fig:Kbb-Fbb_model} and \ref{fig:model_acc}). 
This observed dependence can be fitted with a model  $(w\,f_{\rm c}^{4})\,\ell - w$ using the direct cooling tail method \citep{Suleimanov17}. 
The theoretical models depend on the NS mass $M_{\rm NS}$ and radius $R_{\rm NS}$, which therefore can be constrained for the given X-ray bursting NS. 
The observed data for the X-ray bursts of the investigated source are not good enough for using the direct cooling method. 
Therefore, we fixed the NS mass $M_{\rm NS}=1.5\,M_\odot$ and considered the model curves computed for the fixed surface gravity, $\log g = 14.3$. 
In this case, we obtained two fitting parameters, the observed Eddington flux $F_{\rm Edd}$ and the geometrical dilution factor $\Omega$. 
They depend on the distance to the source $D$, the NS radius $R_{\rm NS}$, and the gravitational redshift $z$ at the NS surface:
\be
    F_{\rm Edd} = \frac{GM_{\rm NS}\,c}{\kappa_{\rm T}} 
    \frac{1}{D^2\,(1+z)}, \qquad \Omega = \frac{R_{\rm NS}^2\,(1+z)^2}{D^2}, 
\ee
where $\kappa_{\rm T} = 0.2(1+X)$\,cm$^2$\,g$^{-1}$ is the electron scattering opacity and $X$ is the hydrogen mass fraction in the atmosphere.

\begin{table}
\caption{Parameters of the NS atmosphere models fitted to the burst spectral evolution.} \label{tab:dr}
\medskip
%\scriptsize
\begin{tabular}{lcccc}
\hline \hline
Parameter & Solar & He (0.0) & He (0.03) & He (0.05) \\
\hline
            \noalign{\smallskip}
$F_{\rm Edd}$ & $\sim$4.28 & $\sim$3.96 & $3.70^{+0.24}_{-0.21}$ & $3.27^{+0.28}_{-0.21}$ \\
            \noalign{\smallskip}
$\Omega$ (km/10\,kpc)$^2$ & $\sim$275 & $\sim$228 & $269^{+4}_{-2}$ & $305^{+10}_{-7}$ \\
            \noalign{\smallskip}
$D$\,(kpc) & $\sim$5.8  & $\sim$8 & $8.2\pm$1.0 & 8.8$^{+1.1}_{-1.5}$\\
            \noalign{\smallskip}
$R_{\rm NS}$\,(km) & $\sim$7.7 & $\sim$9.6 & $10.8\pm$1.1 & 12.3$^{+1.0}_{-1.3}$ \\
\hline 
\end{tabular}\\
\tablefoot{NS mass was assumed  to be $M_{\rm NS}=1.5\,M_\odot$ and gravity $\log g$=14.3. 
The flux is measured in units of  $10^{-8}$\,erg\,s$^{-1}$\,cm$^{-2}$. 
The best-fit distance and the NS radius are obtained for the redshift $z=0.25$.
The errors correspond to the 68.3\% confidence level. 
Numbers in brackets correspond to the accretion luminosity in units of the Eddington one, $L_{\rm acc}/L_{\rm Edd}$. 
\blue{\it }}
\end{table}

We first tried to describe the data by the models computed for two chemical compositions, pure He and solar H/He mix. 
The results are presented in Fig.\,\ref{fig:Kbb-Fbb_model} and the first two columns of Table\,\ref{tab:dr}. 
We note that the obtained distances and the NS radii are rather rough estimations only. 
However, we can conclude that pure He composition of the NS atmosphere is preferred, because the radius estimate is closer to the commonly accepted values of 11--13\,km \citep{2016A&A...591A..25N, 2017A&A...608A..31N, Suleimanov17, 2017MNRAS.472.3905S, 2019PhRvX...9a1001A, 2021ApJ...918L..27R}. 
This conclusion is also in agreement with the low $X$ obtained on the base of $\alpha$ value.

We use only two observational points in the presented above estimations. 
Deviations of other points from the model curves can be explained by the accretion heating \citep{Suleimanov18}. 
Moreover, we can suggest that  accretion heats the NS atmosphere during the whole burst duration, because we observe the pulsations during all burst phases (see Fig.\,\ref{fig:pulse_evol_dur_bursts}).
Therefore, we tried to fit the observed data with the models computed for pure He, $\log g = 14.3$, and two accretion rates corresponding to $L_{\rm acc}/L_{\rm Edd}$ =  0.03  and 0.05. 
The results are presented in Fig.\,\ref{fig:model_acc} and the two rightmost columns of Table\,\ref{tab:dr}. 
These fits give even more  acceptable NS radii of about 11--12\,km.
We note, however, that the cooling tail method is based on an assumption that the NS surface emission is uniform.
This is not correct for the investigated source, because it shows pulsations during the bursts and therefore the obtained estimates are approximate.

\subsection{Modelling the pulse profiles}

We found that the pulse profiles of the persistent emission of \srgash have interesting shapes, following exactly the sine wave in the phase interval 0.0--0.5 and showing a plateau at phases 0.5--1.0 (Fig.~\ref{fig:p_prof_and_PF}). 
There is also some energy dependence in the shape of the plateau. 
At high fluxes, a small bump appears in the middle of the plateau (Fig.~\ref{fig:pulse_prof_5en}). 

%=======================================
\begin{figure}
\centerline{\includegraphics[width=0.8\columnwidth]{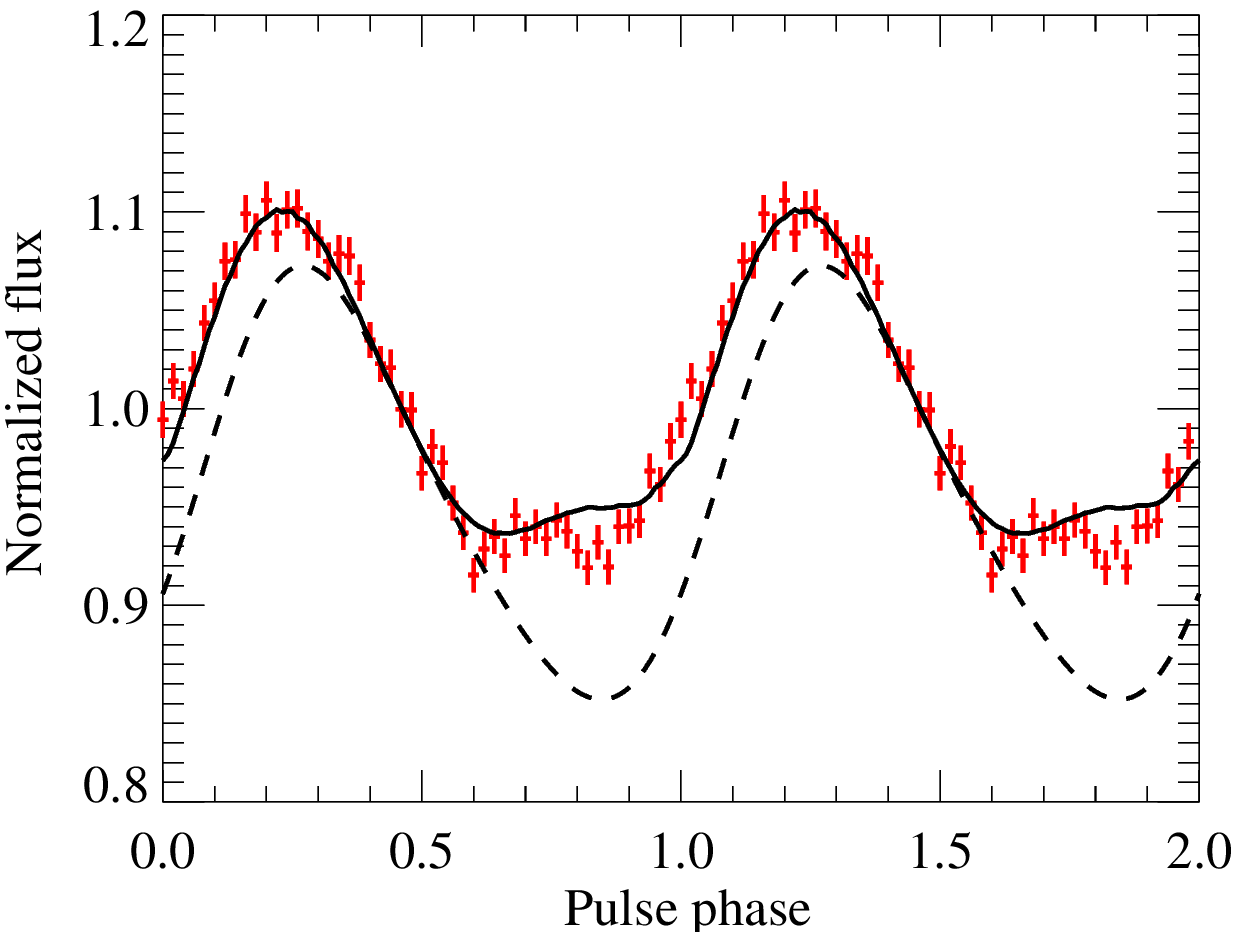}}
\caption{Normalized pulse profiles of the persistent emission of \srgash during the high state (upper data point at Fig.~\ref{fig:p_prof_and_PF}b shown with red crosses) together with an example two-spot model (solid black curve). 
The dashed curve gives the contribution from the primary spot.  }
\label{fig:pp_fit}
\end{figure}
%==============

The simplest explanation for the observed behavior is that two hotspots (associated with the magnetic poles) contribute to the total emission. 
At phases 0.0--0.5, only the primary spot, closest to the observer produces the signal, while at phases 0.5--1.0 there is an additional contribution from the secondary antipodal hotspot. 
Because of the large compactness two antipodal spots can be seen together, and, if their emission pattern is close to the blackbody, two sine waves coming out of phase cancel each other producing a plateau \citep{Beloborodov02,PB06}. 

However, the emission we see in the \art energy range is clearly not a blackbody, but produced by Comptonization in an optically thin region \citep{PG03}. 
Thus, the emission diagram is different and it is not then obvious why the pulse profile has a plateau. 
Alternative explanation involves the effects of partial eclipse of the secondary (southern) hotspot by the  accretion disk \citep{Poutanen09}. 
For a stable accretion to proceed, the inner radius of the accretion disk is expected to lie within the corotation radius, which for a 447.8~Hz pulsar is $R_{\rm co}= 28.6\,(M_{\rm NS}/1.4\,M_\odot)^{1/3}$~km. 
This is just a factor of 2.2 larger than the NS radius. 
Thus, the line of sight towards the southern magnetic pole can pass through the accretion disk. 
The eclipse of the secondary hotspot by the disk was likely responsible for a complex evolution of the pulse profiles during the outburst of SAX~J1808.4$-$3658 \citep{Ibragimov09,Poutanen09,Kajava11}. 

In order to model the pulse profiles, we consider two circular antipodal hotspots at the NS surface, with the primary spot being at co-latitude $\theta$. 
The angular size of the spot is $\rho$,  the observer inclination relative to the NS spin is $i$, and the inner radius of the accretion disk is $R_{\rm in}$. 
The angular emission pattern of the intensity of radiation in the spot frame is parametrized through a linear relation  $I(\mu)\propto 1 + h \cos\alpha'$, where $\alpha'$ is the angle relative to the local normal in the spot comoving frame. 
Parameter $h$ takes the value of 0 for the blackbody emission, while a negative $h$ corresponds to the Comptonized emission from an optically thin slab \citep{PG03,VP04,Bobrikova23}.
For simplicity, we consider Schwarzschild metric and spherical shape of the NS surface. 
We used the code described in \citet{PG03} and \citet{PB06} to model the light curves accounting for disk eclipse following formalism from \citet{Ibragimov09}.  
We fitted by eye the data corresponding to the high state (Fig.~\ref{fig:p_prof_and_PF}b). 
We fixed NS mass to $1.4M_\odot$ and the radius to 12~km. 
We get a good description of the data for the pulsar inclination $i=58\degr$, co-latitude of the spot center of 14\degr\, and its angular radius of 33\degr, the inner disk radius of 24.6~km, the anisotropy parameter $h=-0.63$, and the phase shift of 0.43 (see Fig.~\ref{fig:pp_fit}). 

As we noticed above, the pulse profiles during the bursts (Fig.~\ref{fig:pulse_evol_dur_bursts}) differ substantially from those in the persistent emission. 
This is very different from other bursting AMXPs, e.g., XTE~J1814$-$338 and MAXI~J1816$-$195  \citep{Strohmayer03,Ji24}, which show nearly identical profiles. 
As a result, the same two-spot model with the disk eclipses does not reproduce the data. 
Clearly some model modifications are required. 
%\red{JP: could the reason be that there is frequency evolution during the bursts? }

\section{Summary}
\label{sec:summary}

In this study, we present the results of the analysis of 133~ks of {\it SRG}/ART-XC data, which led to the discovery of a new AMXP, \srgash. Our main results can be summarized as follows: 
\begin{itemize}
%    \item The new accreting millisecond X-ray pulsar \srga\ have been discovered using the {\it SRG}/\art\ data.  
\item The timing analysis revealed a coherent signal near 447.8~Hz modulated by the Doppler effect due to the orbital motion. The derived parameters for the binary system are consistent with the circular orbit with a period of $\sim5.2$~h.    
\item The pulse profiles of the persistent emission, showing a sine-like part during half a period with a plateau in between, can well be modelled by emission from two circular spots partially eclipsed by the accretion disk. 
\item The 19 thermonuclear X-ray bursts have been detected during observations. 
All bursts have similar shapes and energetics and do not show any signs of photospheric radius expansion. The burst rate decreases linearly from one per $\sim$1.6~h at the beginning of observations to one per $\sim$2.2~h at the end and anticorrelates with the persistent flux. 
\item Spectral evolution during the bursts is consistent with the models of NS atmospheres heated by accretion and implies a NS radius of 11--12~km and a distance to the source of 8--9~kpc.
\item The pulsations during the bursts have been detected. We showed the pulse profiles differ substantially from those observed in the persistent emission. 
However, we could not find a simple physical model explaining the pulse profiles detected during the bursts. 

\end{itemize}

\begin{acknowledgements}

This work is based on observations with the {\it Mikhail Pavlinsky} \art\ telescope, hard X-ray instrument on board the \srg\  observatory. The \srg\ observatory was created by Roskosmos  in the interests of the Russian Academy of Sciences represented by its Space Research Institute (IKI) in the framework of the Russian Federal Space Program, with the participation of Germany. The ART-XC team thanks Lavochkin Association (NPOL) with partners for the creation and operation of the SRG spacecraft (Navigator). This work was supported by the grant Minobrnauki 23-075-67362-1-0409-000105. 
SST and JP acknowledge the Academy of Finland grants 333112, 349373,  and 349906. 
VFS thank the Deutsche Forschungsgemeinschaft (DFG) for financial support (grant
WE 1312/59-1).

\end{acknowledgements} 

\bibliographystyle{yahapj}
\bibliography{biblio.bib}

\end{document}